# Proton irradiation on Hydrogenated Amorphous Silicon flexible devices

M. Menichelli[1,*], S.Aziz[2], A. Bashiri[3,4], M.Bizzarri[1,5], C. Buti[6,7], L.Calcagnile[2], D. Calvo[8], M. Caprai[1], D. Caputo[9,10], A.P. Caricato[2], R. Catalano[11], M. Cazzanelli[12], R. Cirio[8], G.A.P. Cirrone[11], F. Cittadini[1,22], T.Croci[1,13], G. Cuttone[11], G. de Cesare[9,10], P. De Remigis[8], S.Dunand[14], M.Fabi[15,6], L.Frontini[16,17], C.Grimani[15,6], M. Guarrera[11], H.Hasnaoui[12], M. Ionica[1], K. Kanxheri[1,5], M. Large[3], F. Lenta[8,18], V.Liberali[16,17], N.Lovecchio[9,10], M.Martino[2], G. Maruccio[2], L. Maruccio[2], G.Mazza[8], A. G. Monteduro[2], A. Morozzi[1], A. Nascetti[9,19], S. Pallotta[6,7], A. Papi[1], D. Passeri[1,13] *Senior Member IEEE*, M.Pedio[1,20], M. Petasecca[3] *Member IEEE*, G.Petringa[11], F.Peverini[1,5], P.Placidi[1,13] *Senior Member IEEE*, M.Polo[12], A. Quaranta[12], G.Quarta[2], S. Rizzato[2], F.Sabbatini[6], L. Servoli[1], A. Stabile[16,17], C. Talamonti[6,7], J. E.Thomet[14], M.S. Vasquez Mora[16], M.Villani[15,6], R.J. Wheadon[8], N. Wyrsch[14], N.Zema[1,21] and L. Tosti[1*].

[1]The HASPIDE project is funded by INFN through the CSN5 and was partially supported by the "Fondazione Cassa di Risparmio di Perugia" RISAI project n. 2019.0245. F. Peverini has a PhD scholarship funded by the Programma Operativo Nazionale (PON) program. M. J. Large is supported by the Australian Government Research Training Program (AGRTP) scholarship and the Australian Institute of Nuclear Science (AINSE) Post-Graduate Research Award (PGRA). A. Bashiri is sponsored by Najran University, Saudi Arabia. J. E. Thomet are supported by the Swiss National Science Foundation (grant number 200021_212208/1).
  This paper was presented at the IEEE-RTSD2024 in Tampa (USA) with presentation number: R08-04

[1.] INFN, Sez. di Perugia, via Pascoli s.n.c. 06123 Perugia (ITALY)
[2.]INFN and Dipartimento di Fisica e Matematica dell'Università del Salento, Via per Arnesano, 73100 Lecce (ITALY)
[3.] Centre for Medical Radiation Physics, University of Wollongong, Northfields Ave Wollongong NSW 2522, (AUSTRALIA)
[4.]Najran University, King Abdulaziz Rd, Najran. (Saudi Arabia)
[5.] Dip. di Fisica e Geologia dell'Università degli Studi di Perugia, via Pascoli s.n.c. 06123 Perugia (ITALY)
[6.]INFN Sez. di Firenze, Via Sansone 1, 50019 Sesto Fiorentino (FI) (ITALY)
[7.] Department of Experimental and Clinical Biomedical Sciences "Mario Serio"- University of Florence Viale Morgagni 50, 50135 Firenze (FI) (ITALY)
[8.]INFN Sez. di Torino Via Pietro Giuria, 110125 Torino (ITALY)
[9.]INFN Sezione di Roma 1, Piazzale Aldo Moro 2, Roma (ITALY)
[10.]Dipartimento di Ingegneria dell'Informazione, Elettronica e Telecomunicazioni, dell'Università degli studi di Roma via Eudossiana, 18 00184 Roma (ITALY).
[11.]INFN Laboratori Nazionali del Sud, Via S.Sofia 62, 95123 Catania (ITALY)
[12.]TIFPA and Trento University, Via Sommarive 14, 38123 Povo, TN (Italy)
[13.]Dip. di Ingegneria dell'Università degli studi di Perugia, via G.Duranti 06125 Perugia (ITALY)
[14.]Ecole Polytechnique Fédérale de Lausanne (EPFL), Institute of Electrical and Microengineering (IME), Rue de la Maladière 71b, 2000 Neuchâtel, (SWITZERLAND).
[15.]DiSPeA, Università di Urbino Carlo Bo, 61029 Urbino (PU) (ITALY)
[16.]INFN Sezione di Milano Via Celoria 16, 20133 Milano (ITALY)
[17.]Dipartimento di Fisica dell'Università degli Studi di Milano, via Celoria 16. 20133 Milano (ITALY).
[18.] Politecnico di Torino, Corso Duca degli Abruzzi 24, 10129 Torino, (ITALY).
[19.]Scuola di Ingegneria Aerospaziale Università degli studi di Roma.Via Salaria 851/881, 00138 Roma. (ITALY).
[20.]CNR-IOM, via Pascoli s.n.c. 06123 Perugia (ITALY)
[21.]CNR Istituto struttura della Materia, Via Fosso del Cavaliere 100, Roma (ITALY).
[22] Dip. Di Fisica e Astronomia dell'Università di Padova, via Marzolo 8, 35131 Padova (ITALY)

[1]*Abstract*— **Radiation damage tests in hydrogenated amorphous silicon (a-Si:H) flexible flux and dose-measuring devices have been performed with a 3 MeV proton beam, to evaluate combined displacement and total ionizing dose damage. The tested devices had two different configurations and thicknesses. The first device was a 2 μm thick n-i-p diode having a 5 mm x 5 mm area. The second device was a 5 μm thick charge selective contact detector having the same area. Both the devices were deposited on a flexible polyimide substrate and were irradiated up to the fluence of $10^{16}$ $n_{eq}/cm^2$. The response to different proton fluxes has been measured before irradiation and after irradiation at $10^{16}$ $n_{eq}/cm^2$ for charge-selective contacts and n-i-p devices. The effect of annealing for partial performance recovery at 100°C for 12 hours was also studied and a final characterization on annealed devices was performed. This test is the first combined displacement and total ionizing dose test on flexible a-Si:H devices.**

## I. Introduction

HYDROGENATED Amorphous Silicon (a-Si:H) devices on a flexible substrate were developed in the framework of the HASPIDE project aiming at the fabrication of these devices for dosimetry, beam flux measurement, neutron flux measurement and space applications [1],[2],[3]. A-Si:H is a disordered semiconductor where the Si-Si bonds have different lengths and where unsaturated bonds, also called dangling bonds (DBs), are present. The resulting structure is irregular with some voids inside, for this reason, the density of this material is lower than crystalline Silicon (2.33 for crystalline Silicon and 2.285 for a-Si:H) [4]. The band structure of this material has two band-tails, one in the valence band ($\Delta E_V$ = 40-60 meV) and the other in the conduction band ($\Delta E_C$ = 20-40 meV); the presence of DBs generates mid-band defects that act as traps for the ionization charge generated by the radiations inside the material reducing charge collection efficiency.

The very high level of radiation hardness of a-Si:H was proven in the early period of a-Si:H application for radiation detectors [5]. For this property, these detectors gained very much attention in the field of high-energy physics experiments and also for medical imaging [6]. Several radiation damage tests were performed using protons [7-10], gammas [11], neutrons [12], or heavy ions irradiation [13] on p-i-n devices. A very important proton irradiation test was performed at CERN on 32 µm thick a-Si:H diodes to study defect formation and metastability. These a-Si:H devices were exposed to a 24 GeV proton beam in the 'IRRAD1' facility up to fluencies of $2 \times 10^{16}$ protons/cm$^2$ [14]. From the data shown in the reference, it is possible to observe a slow degradation of the detector response (namely a decrease of the current response) at increasing fluences which is observed to saturate above a fluence of $10^{15}$ protons/cm$^2$ at a value which the current response is around 50% the original value. The self-annealing effect, related to the metastable nature of the material, was also observed when the irradiation was stopped for 20 h.

* Corresponding author, Mauro.menichelli@pg.infn.it, luca.tosti@pg.infn.it

Radiation resistance for high-energy physics applications has also been studied, with neutron irradiation, by this collaboration [15]. The comparison between the CERN experiment and the neutron irradiation, for a neutron equivalent fluence up to $10^{16}$ n$_{eq}$/cm$^2$, gave similar results showing an increment of a factor 2 in leakage current and a loss in sensitivity. These degradations were reversed after 12 hours of annealing at 100 °C.

The deposition of a-Si:H on a thin and flexible substrate (like polyimide) allows the fabrication of thin and flexible devices opening the possibility of usage where flexibility, transparency to radiation and radiation resistance are important factors in applications like those mentioned at the beginning of this section.

In this paper, the results of an irradiation test with protons, combining displacement damage and total ionization damage, will be shown and compared with an irradiation test with neutrons and a photo-emission spectroscopy study previously performed at ELECTRA Synchrotron in Trieste on neutron irradiated p-i-n devices.

## II. THE TEST FACILITY AT CEDAD, THE TEST SETUP AND THE DETECTORS

This test has been performed at the CEDAD irradiation facility of the University of Salento in Brindisi (Italy) using a 3 MeV proton beam. The accelerator is a tandem providing protons or ions (not used for this test) with a maximum energy (for protons) of 5 MeV up to very high fluxes (above 1 µA). The detectors under test were placed in a vacuum chamber connected to a beamline (Fig.1). Inside this chamber, a Faraday cup (FC) is located where the beam was dumped. This FC was used to measure the beam flux before placing the detector on the beam and in "transparency mode" with the detector placed in front of the FC, which can be used to monitor the stability of the flux during irradiation and testing.

The detector response was calibrated by measuring the current response versus proton fluxes; the beam spot size was circular with a 1 cm diameter. The flux was measured by the FC and then normalized to the area of the spot. The precision of this measurement is estimated at around ±10% from the previous evaluation.

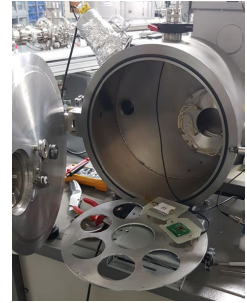

Fig.1. Irradiation chamber of the accelerator at CEDAD. The beam enters from the flange on the right inside the chamber. The picture also shows a test stand where a detector and a ceramic target to evaluate the beam spot size are placed. On the left in the door of the chamber, there is a flange where the Faraday cup is located (not shown).

Two different types of detectors were tested (Fig.2): a n-i-p detector [2] having an intrinsic a-Si:H sensitive thickness of 2 µm and pad area of 5 mm x 5 mm and a charge selective contacts (CSC) detector [16] having an intrinsic a-Si:H sensitive thickness of 5 µm and same pad area (Fig.3 and 4).

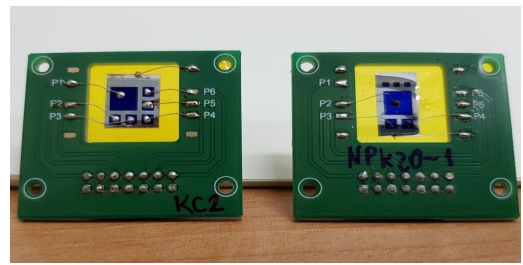

Fig.2 On the left the CSC detector having 5 µm intrinsic a-Si:H thickness is shown. It has one 5 mm x 5 mm pad and five 2 mm x 2 mm pads but only the 5 mm x 5 mm have been characterized in this test. On the right a n-i-p detector having 2 µm intrinsic a-Si:H thickness is displayed. It includes one 5 mm x 5 mm pad, two 2 mm x 2 mm pads and three 1 mm x 1 mm pads (not bonded), also in this case only the large pad has been characterized. Both detectors are mounted on a PCB frame to make the contacts with a readout connector.

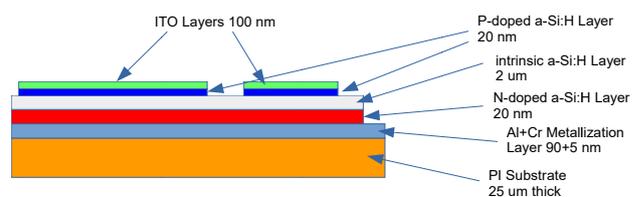

Fig.3 Structure of a n-i-p detector. The thickness of the intrinsic a-Si:H is 2 µm and the detector pad had a 5 mm x 5 mm area.

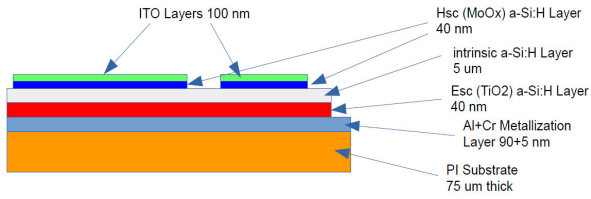

Fig.4 Structure of a CSC device. ESC stands for electron-selective contact (Titanium Oxide), and HSC stands for hole-selective contact (Molybdenum Oxide).

Detectors were characterized before irradiation, after $10^{16}$ $n_{eq}/cm^2$ irradiation (corresponding to 1.20 Grad of Total Ionizing Dose), and after annealing at 100°C lasting 12 hours. During the measurements, the devices current was measured with a Keithley Source Measurement Unit (2400 model) with a resolution of 10 pA.

### III. IRRADIATION TEST AND PRELIMINARY RESULTS

The two devices described above were characterized using the same proton beam used for the radiation damage test with fluxes ranging from $10^9$ p/(cm² *s) to $3 * 10^{10}$ p/(cm² *s) both under bias (-1V for the n-i-p device and -2V for the CSC device) and at 0 V bias. During the characterization procedure, the current was measured at different proton fluxes in the range above mentioned and the background current was taken as an average between the current before and after the irradiation with the proton flux used for characterization, during the data analysis the background current has been subtracted to the current measured during irradiation. After this initial characterization, radiation damage-oriented irradiation at $10^{11}$ p/(cm² *s) was performed up to the total neutron equivalent dose of $10^{16}$ $n_{eq}/cm^2$, the components were irradiated under bias. After the irradiation, a characterization identical to the one described above was performed. After this second characterization, annealing at 100 °C for 12 h followed by a new characterization test with the proton beam was performed.

The results for the n-i-p device are displayed in Fig.5 at 0V and in Fig.6 at -1 V bias. It can be observed that after the $10^{16}$ $n_{eq}/cm^2$ irradiation, the sensitivity at 0V (namely the slope of the line fitting the current versus the proton flux) drops by more than an order of magnitude (7% of the original value), while after the annealing, this parameter recovers up to 59.6% of its value before irradiation, concerning the regression coefficient R of the fit, a very good linearity was observed. For the biased detector the sensitivity degradation after the irradiation (13% of the original value) was lower, and a better sensitivity recovery after annealing, which even reached a higher value than before the irradiation (107%), was observed.

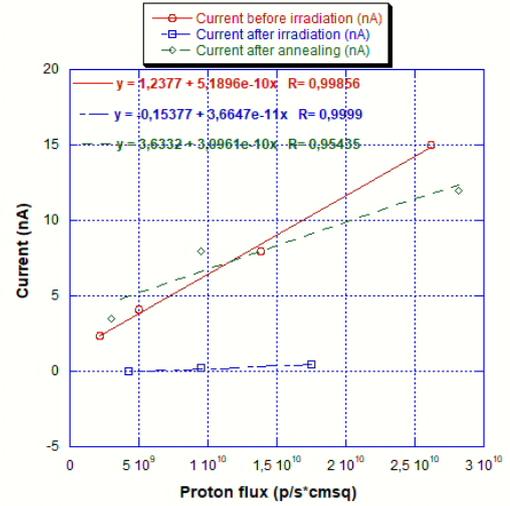

Fig. 5. Measured n-i-p device current values (nA) at 0V bias versus proton flux (p/(cm² * s)) before and after irradiation, and after annealing. The slope of the three lines represents the sensitivity.

Concerning the CSC device at 0V bias (Fig.7), also in this case, after a drop of more than one order of magnitude in sensitivity after irradiation, a recovery is observed (up to 53% of the original value) after annealing.

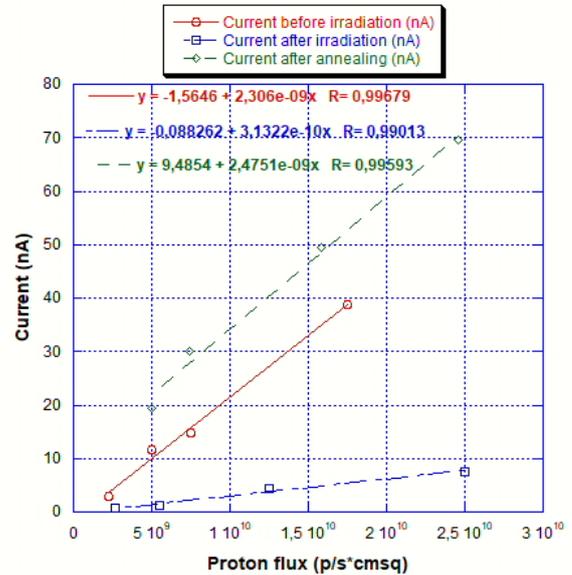

Fig. 6. Measured n-i-p device current values (nA) at -1V bias versus proton flux (p/(cm² * s)) before and after irradiation, and after annealing.

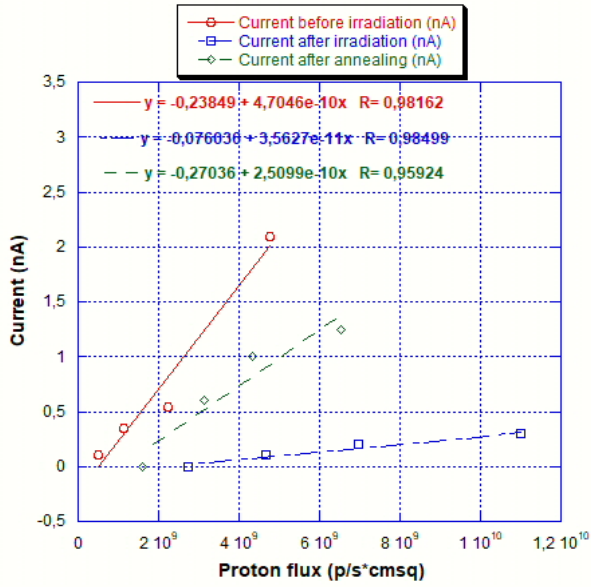

Fig. 7. Measured CSC device current values (nA) at 0 V bias versus proton flux (p/cm$^2$ * s) before and after irradiation and after annealing.

At -2V bias the sensor shows the usual decrease of one order of magnitude in sensitivity (7% of the original value) after irradiation and a recovery of up to 60% after annealing. Table I shows a summary of all measurements.

In summary, it is observed a degradation of one order of magnitude in sensitivity after irradiation for both types of detector, a complete recovery for n-i-p detectors under biased condition and a partial recovery of about 50-60% on n-i-p unbiased. For CSC a partial recovery both for biased and unbiased sensors was observed after annealing. The linearity remains at an acceptable level throughout all measurements.

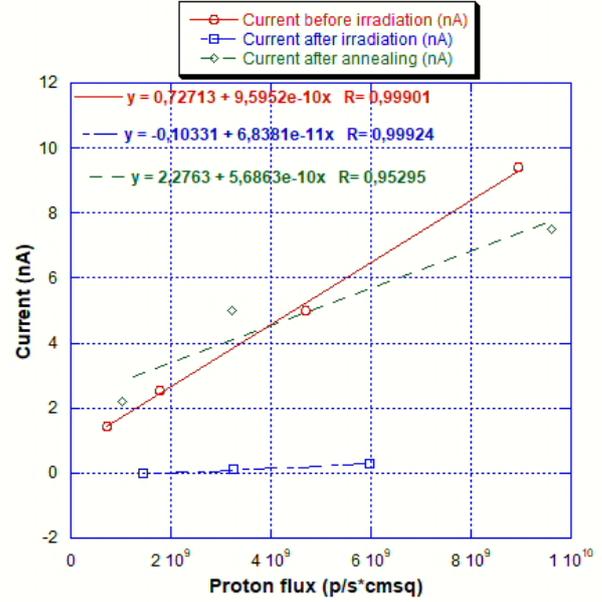

Fig.8. Measured CSC device current values (nA) at - 2 V bias versus proton flux (p/cm$^2$ * s) before and after irradiation and after annealing.

TABLE I

| Summary of measurements<br>*Samples and measurements* | *Sensitivity*<br>*(nC·cm$^2$·10$^{10}$)* | *% of sensitivity compared to an unirradiated sample* | *R* |
|---|---|---|---|
| n-i-p at 0V bias before irradiation | 5.1896 | 100% | 0.9986 |
| n-i-p at -1V bias before irradiation | 23.06 | 100% | 0.997 |
| n-i-p at 0V bias after irradiation | 0.366 | 7.0% | 1 |
| n-i-p at -1V bias after irradiation | 3.13 | 13.6% | 0.9901 |
| n-i-p at 0V bias after annealing | 3.0961 | 59.6% | 0.9543 |
| n-i-p at -1V bias after annealing | 24.75 | 107.3% | 0.9959 |
| CSC at 0V bias before irradiation | 4.705 | 100% | 0.9816 |
| CSC at -2 V bias before irradiation | 9.595 | 100% | 0.9990 |
| CSC at 0V bias after irradiation | 0.3563 | 7.6% | 0.9850 |
| CSC at -2V bias after irradiation | 0.6838 | 7.1% | 0.9992 |
| CSC at 0V bias after annealing | 2.510 | 53.3% | 0.9592 |
| CSC at -2V bias after annealing | 5.686 | 59.3% | 0.9529 |

## IV. COMPARISON WITH PREVIOUS TESTS AND DISCUSSION

The results of this proton irradiation of flexible a-Si:H radiation sensors can be compared with a neutron irradiation test on a-Si:H sensors on different substrates and with different thicknesses (p-i-n 10 μm thick a-Si:H layer and CSC with 8.2 μm thick a-Si:H layer) and operating condition (irradiation unbiased). The test on these detectors was performed in the framework of the 3D-SiAm experiment [17] at the same equivalent fluence (i.e. $10^{16} n_{eq}$ /cm$^2$), the test is described in detail in reference [15]. The characterization of these detectors was performed using X-rays and the test protocol was similar to the one used for the experiment described here, namely:

- Initial characterization.
- Irradiation up to the fluence of $10^{16} n_{eq}$ /cm$^2$ .
- Post-irradiation characterization
- Annealing of 12 hours at 100°C
- Final characterization.

The variation of sensitivities for the various phases of the two tests are shown in Table II, although a comparison has been performed between X-ray sensitivities and proton flux sensitivities.

TABLE II
COMPARISON BETWEEN NEUTRON TEST AND PROTON TEST SENSITIVITY DEGRADATION

| Detector type and condition | Neutron test sensitivity %compared to pre-irradiation | Proton test sensitivity% compared to pre-irradiation |
|---|---|---|
| n-i-p (or p-i-n) device biased after irradiation | 47.3% | 13.6% |
| CSC device at 0 V bias after irradiation | 11.6% | 7.6% |
| CSC device biased after irradiation | 45.0% | 7.1% |
| n-i-p (or p-i-n) device biased after annealing | 125.5% | 107.3% |
| CSC device at 0 V bias after annealing | 27.9% | 53.3% |
| CSC device biased after annealing | 58.6% | 59.3% |

It is possible to notice that the qualitative behavior is similar for the two different tests: there is a sensitivity degradation after irradiation and a partial recovery after annealing in CSC, and a total recovery in n-i-p/p-i-n devices, in both cases (proton and neutron irradiations). From the quantitative point of view, the sensitivity loss is larger with the proton irradiation, this is probably due to the combined effect of displacement damage and ionization damage, in any case, the sensitivity loss is smaller if the characterization is performed under bias. Actually, in the neutron test, where the sensitivity loss is smaller, the p-i-n was biased with a field of 6 V/cm while in the proton test, the biasing field was only 0.5 V/cm and for CSC in the neutron test the biasing field was 3.49 V/cm in neutron irradiation, compared with the 0.4 V/cm for the proton irradiation. After annealing it is observed a complete recovery of the damage in n-i-p/p-i-n devices both for proton and for neutron irradiations. For CSC devices a partial recovery, of about the same amount for the two types of irradiation is observed in biased devices, while a better recovery for neutron irradiation is observed in biased CSC devices compared to unbiased.

Since the present data were taken in different conditions, to have a more precise picture of the irradiation results, the collaboration will perform a neutron irradiation test using similar devices as in the proton test.

To improve the knowledge of the radiation damage mechanism at the structural level in a-Si:H a photo emission spectroscopy test, on a few irradiated samples with neutrons [18] has been performed. The goal of the measurement was the estimation of the percentage of the Si-H bonds contribution and of the disordered component including the dangling bonds that generate defects. These defects generate traps that lower the charge collection efficiency and may be saturated by the Si-H bonds. The most relevant issue is to monitor how these quantities change after irradiation and subsequent annealing, additional information about the presence of polycrystalline silicon structures in the amorphous silicon and the bonds with Carbon impurities are also provided from this test. The results of these studies are summarized in Table III.

TABLE III
PERCENTAGE OF BOND TYPES IN THE REFERENCE SAMPLE AND THE DETECTOR AT THE VARIOUS STAGES OF THE IRRADIATION PROCEDURE

| | Reference Sample | Before Irradiation | After Irradiation | After Annealing |
|---|---|---|---|---|
| c-Si poly-crystalline | 4% | 38% | 17% | 21% |
| Disordered and angling bonds | 70% | 46% | 67% | 44% |
| Si-H Bonds | 15% | 11% | 3% | 12% |
| Si-C Bonds | 11% | 5% | 13% | 23% |

The reference sample is just a simple a-Si:H layer while the other column refers to an actual a-Si:H p-i-n detector deposited on Silicon. The percentage of poly-crystalline Si is much higher in a detector compared to the bare a-Si:H reference sample and this depends on the various fabrication stages, that generate local warming-up of the material. This causes the formation of polycrystalline "domains" on the amorphous structure of the material. During the irradiation, these structures became smaller and have an increase after annealing but the percentage remains smaller compared to the non-irradiated sample. Concerning dangling bonds, they increase after irradiation due to displacement damage and decrease back to the original value after annealing. Si-H bonds, which usually passivate the dangling bonds, tend to decrease after irradiation causing the hydrogen to become an interstitial impurity; after annealing the hydrogen bonds with Silicon are restored up to the original value. Si-C bonds generated by carbon impurities in a-Si:H, keep increasing during the annealing process.

According to the results from photoemission and its interpretation, displacement damage induced by irradiation is due to the breaking of Si-Si and Si-H bonds that can be reversed with annealing. A further spectroscopic study on proton irradiated detectors deposited on polyimide is important to verify these findings.

## V. Conclusions

Proton damage with fluencies at the level of $10^{16}$ $n_{eq}$ /cm$^2$ displacement dose and 1.2 Grad total ionizing dose have been studied on flexible a-Si:H sensors (having n-i-p and CSC configuration) at the CEDAD proton accelerator using 3 MeV protons. After irradiation, a subsequent annealing at 100°C degrees for 12 hours has been performed. The components under test were irradiated up to the target fluence under bias and the characterization measurement of sensitivity was performed both at 0V bias and biased (-1V for the n-i-p detector and -2 V for the CSC detector). After irradiation, the results show a reduction of one order of magnitude in sensitivity for n-i-p and CSC detectors. This reduction is fully recovered in the n-i-p device after the annealing and partially recovered in CSC devices. This result is in qualitative agreement with a neutron irradiation test at the same displacement dose on samples deposited on different substrates. A photo-emission spectroscopy study on a p-i-n device irradiated with neutron shows that the damage in terms of loss of charge collection efficiency is due to an increase of defects in the material and a decrease of Si-H bonds that can be recovered with annealing.